\documentstyle[times,verbatim,twocolumn]{article}

\makeatletter
\def\@normalsize{\@setsize\normalsize{10pt}\xpt\@xpt
\abovedisplayskip 10pt plus2pt minus5pt\belowdisplayskip \abovedisplayskip
\abovedisplayshortskip \z@ plus3pt\belowdisplayshortskip 6pt plus3pt
minus3pt\let\@listi\@listI}
\def\subsize{\@setsize\subsize{12pt}\xipt\@xipt}
\def\section{\@startsection {section}{1}{\z@}{1.0ex plus 1ex minus
 .2ex}{.2ex plus .2ex}{\large\bf}}
\def\subsection{\@startsection {subsection}{2}{\z@}{.2ex plus 1ex}
{.2ex plus .2ex}{\subsize\bf}}
\makeatother

\newtheorem{theorem}{\sc Theorem}
\newtheorem{lemma}{\sc Lemma}
\newtheorem{coro}{\sc Corollary}
\newtheorem{nota}{\sc Notation}
\newtheorem{rem}{\sc Remark}
\newtheorem{defin}{\sc Definition}
\newtheorem{cla}{\sc Claim}
\newtheorem{ex}{\sc Example}
\newenvironment{proof}{\par \sc Proof.\rm}{\hspace*{\fill}$\Box$\vspace{1ex}}
\newenvironment{example}{\begin{ex}}{\hspace*{\fill}$\Diamond$\end{ex}}

\newenvironment{corollary}{\begin{coro}}{\end{coro}}
\newenvironment{definition}{\begin{defin}}{\end{defin}}

\newenvironment{remark}{\begin{rem}}{\hspace*{\fill}$\Diamond$\end{rem}}
\newcommand{\trace}[1]{\mbox{Tr$( #1 )$}}
\newcommand{\ket}[1]{\mbox{$| #1 \rangle$}}

\newcommand{\bracket}[2]{\mbox{$\langle #1|#2 \rangle$}}
\newcommand{\shar}[2]{\mbox{$ | #1 \rangle \langle #2 |$}}

\begin{document}

\title{Three Approaches to the
 Quantitative Definition of Information in an Individual Pure
Quantum State}
\author{Paul Vitanyi\thanks{Partially supported by the EU fifth framework
project QAIP, IST--1999--11234, the NoE QUIPROCONE IST--1999--29064,
the ESF QiT Programmme, and ESPRIT BRA IV NeuroCOLT II Working Group
EP 27150.
Part of this work was done
during the author's 1998 stay at Tokyo Institute of Technology,
Tokyo, Japan, as Gaikoku-Jin Kenkyuin at INCOCSAT. A preliminary version 
was archived as quant-ph/9907035. Address:
CWI, Kruislaan 413, 1098 SJ Amsterdam, The Netherlands. Email:
{\tt paulv@cwi.nl}}\\
CWI and University of Amsterdam}
\date{}
\maketitle
\thispagestyle{empty}

\begin{abstract}
In analogy of classical Kolmogorov complexity
we develop a theory of the algorithmic information in bits contained
in any one of continuously many pure quantum states:
quantum Kolmogorov complexity.
Classical Kolmogorov complexity
coincides with the new quantum Kolmogorov complexity restricted to
the classical domain. Quantum Kolmogorov complexity is upper
bounded and can be effectively approximated from above.
With high probability a quantum object is incompressible. 
There are two alternative approaches possible:
to define the complexity as the length of the shortest
qubit program that effectively describes the object, and to
use classical descriptions with
computable real parameters.
% We regard these approaches as less
%promising for future applications and relegate them to the appendix.
%This is a preliminary version---a more detailed version follows.
\end{abstract}
\section{Introduction}
While Kolmogorov complexity is the accepted absolute measure
of information content in a {\em classical} individual finite object,
a similar absolute notion is needed for the information
content of a pure quantum state.
\footnote{
For definitions and theory of Kolmogorov complexity consult \cite{LiVi97},
and for quantum theory consult \cite{Pe95}.
}
Quantum theory assumes that every complex vector, except the null vector,
represents a realizable pure quantum state.\footnote{That is,
every complex vector that can be normalized to unit length.}
 This leaves open the question
of how to design the equipment that prepares such a pure
state. While there are continuously many pure states in 
a finite-dimensional complex vector space---corresponding to all vectors of
unit length---we can finitely
describe only a countable subset. Imposing effectiveness on
such descriptions leads to constructive procedures.
The most general such procedures satisfying universally agreed-upon
logical principles of effectiveness are quantum Turing machines, \cite{BV97}.
To define quantum Kolmogorov complexity 
by way of quantum Turing machines leaves
essentially two options:
\begin{enumerate}
\item
We want to describe every quantum superposition exactly; or
\item
we want to take
into account the number of bits/qubits in the specification
as well the accuracy of the quantum state produced. 
\end{enumerate}
We have to deal with three problems:
\begin{itemize}
\item
There are continuously many quantum Turing machines;
\item
There are continously many pure quantum states;
\item
There are continuously many qubit descriptions.
\end{itemize}
There are uncountably many quantum Turing machines
only if we allow arbitrary real rotations in the definition of 
machines. Then, a quantum Turing machine can only be universal
in the sense that it can approximate the computation of an
arbitrary machine, \cite{BV97}. In descriptions using universal
quantum Turing machines
we would have to account for the closeness of approximation,
the number of steps required to get this precision, and the like.
In contrast, if we fix the rotation
of all contemplated machines to a single primitive rotation $\theta$ 
with $\cos \theta = 3/5$ and $\sin \theta = 4/5$ then there
are only countably many Turing machines and the universal machine
simulates the others exactly \cite{ADH97}. 
Every quantum
Turing machine computation using arbitrary real rotations
can be approximated to any precision by machines with fixed
rotation $\theta$ but in general cannot be simulated 
exactly---just like in the case of the simulation of
arbitrary quantum Turing machines by a universal
quantum Turing machine.  Since exact simulation is impossible
by a fixed universal quantum Turing machine anyhow, but arbitrarily
close approximations are possible by Turing machines using
a fixed rotation like $\theta$, we are motivated to fix 
$Q_1 , Q_2 , \ldots$ as a standard enumeration of
quantum Turing machines using only rotation $\theta$.

Our next question is whether we want programs 
(descriptions) to be in classical bits
or in qubits?
The intuitive notion of computability requires
the programs to be classical. Namely, to prepare a quantum state
requires a physical apparatus that ``computes'' this quantum state
from classical specifications. 
Since such specifications have effective descriptions,
every quantum state that can be prepared can 
be described effectively in descriptions consisting of classical bits.
Descriptions consisting of arbitrary pure quantum states
allows noncomputable (or hard to compute)
information to be hidden in the bits of
the amplitudes.
In Definition~\ref{def.pqscomp} we call a pure quantum state {\em directly
computable} if there is a (classical) program such that
the universal quantum Turing machine computes that state from
the program and then halts in an appropriate fashion.
In a computational setting we naturally
require that directly computable pure quantum states can be 
prepared.
By repeating the preparation we can obtain
arbitrarily many copies of the pure quantum state.
\footnote{See the discussion in \cite{Pe95}, pp. 49--51.
If descriptions are not effective then we are not going to use them in our
algorithms except possibly on inputs from an ``unprepared''
origin. Every quantum state used in a quantum computation
arises from some classically preparation or is possibly
captured from some unknown origin. If the latter, then we can consume
it as conditional side-information or an oracle.} 
Restricting ourselves to an effective enumeration of
quantum Turing machines and classical descriptions
to describe by approximation continuously many pure quantum states is
reminiscent of the construction of continuously many real  numbers
from Cauchy sequences of rational numbers, the rationals being 
effectively enumerable. 

The second approach 
considers the shortest effective qubit description
of a pure quantum state. This can also be properly
formulated in terms of the conditional version
of the first approach. An advantage of this version is that the
upper bound on the complexity of a pure
quantum state is immediately given by the number of qubits involved in the
literal description of that pure quantum state. 
The status
of incompressibility and degree of uncomputability is as yet
unknown and potentially a source of problems with this approach.

The third approach is to give programs for the $2^{n+1}$ real numbers
involved in the precise description of the $n$-qubit state. Then
the question reduces to the problem of describing lists of 
real numbers.

In the classical situation there are also several variants
of Kolmogorov complexity that are very meaningful in their
respective settings: plain Kolmogorov complexity, prefix complexity,
monotone complexity, uniform complexity,
negative logarithm of universal measure, and so on \cite{LiVi97}.
It is therefore not surprising that in the more complicated situation
of quantum information several different choices of complexity 
can be meaningful and unavoidable in different settings. 

\section{Classical Descriptions}

The complex
quantity $\bracket{x}{z}$ is the inner product of vectors $\ket{x}$ 
and $\ket{z}$.
Since pure quantum states $\ket{x}, \ket{z}$ have unit length, 
$|\bracket{x}{z}| =  | \cos \theta |$  where $\theta$ is the
angle between vectors $\ket{x}$ and $\ket{z}$
and $|\bracket{x}{z}|^2$ is the probability of outcome
$\ket{x}$ being
measured from state $\ket{z}$, \cite{Pe95}.  The idea is as follows.
A {\em von Neumann measurement} is a decomposition of the Hilbert space 
into subspaces that are mutually orthogonal, for example an 
orthonormal basis is an observable. Physicists like to 
specify observables as Hermitian matrices, where the 
understanding is that the eigenspaces of the matrices 
(which will always be orthogonal) are the actual subspaces.
When a measurement is performed, the state is projected 
into one of the subspaces (with probability equal to the square of 
the projection). So the subspaces correspond to the possible {\em outcomes} 
of a measurement. In the above case we project $\ket{z}$ on outcome
$\ket{x}$ using projection $\shar{x}{x}$ resulting in
$\bracket{x}{z} \ket{x}$.

Our model of computation is a quantum
Turing machine with classical binary program $p$ on the input tape
and a quantum auxiliary input 
on a special conditional input facility. We think of this auxiliary input
as being given as a pure quantum state $\ket{y}$ 
(in which case it can be used only once),
as a mixture density matrix $\rho$, or
(perhaps partially) as a classical program from which
it can be computed. In the last case, the classical program can of course
be used indefinitely often.\footnote{We can even allow that the conditional
information $y$ is infinite or noncomputable, or an oracle. 
 But we will not need this in the present paper.
}
It is therefore not only important {\em what} information is
given conditionally, but also {\em how} it is described---like
this is the sometimes the case in the classical version
of Kolmogorov complexity for other reasons that would additionally
hold in the quantum case.
 We impose the condition that the set of {\em halting
programs} ${\cal P}_y  = \{p: T(p | y) < \infty \}$ is {\em prefix-free}:
no program in ${\cal P}_y$ is a proper prefix of another program in
${\cal P}_y$. Put differently, the Turing machine scans all of a
halting program $p$ but never scans the bit following the last
bit of $p$: it is {\em self-delimiting}.
\footnote{One can also use a model were the input $p$ is delimited
by distinguished markers. Then the Turing machine always knows where
the input ends. In the self-delimiting case the endmarker must be
implicit in the halting program $p$ itself. This encoding of the
endmarker carries an inherent penalty in the form of increased length:
typically a prefix code of an $n$-length binary string has length
about $n+ \log n + 2 \log \log n$ bits, \cite{LiVi97}.}
\footnote{
There are two possible interpretations for the computation relation
$Q(p, y) = \ket{x}$. In the narrow interpretation
we require that $Q$ with $p$ on the input tape
and $y$ on the conditional tape halts with $\ket{x}$
on the output tape. In the wide interpretation we can
define pure quantum states by requiring
that for every precision $\delta > 0$ the computation
of $Q$ with $p$ on the input tape
and $y$ on the conditional tape
and $\delta$ on a tape where the precision is to be supplied
halts with $\ket{x'}$
on the output tape and $|\bracket{x}{x'}|^2 \geq 1-\delta$.
Such a notion of ``computable''
or ``recursive'' pure quantum states is similar to Turing's
notion of ``computable numbers.'' 
In the remainder of this section we use the narrow interpretation.
}

\begin{definition}
\rm
The {\em (self-delimiting) complexity} of $\ket{x}$ 
with respect to quantum Turing machine $Q$
with $y$ as conditional input given for free is 
\[
K_Q (\ket{x} | y ) := 
\min_{p} \{ l(p) + \lceil - \log(|\bracket{z}{x}|^2) \rceil : 
Q(p, y) = \ket{z} \}
\]
where $l(p)$ is the number of bits in the specification $p$,
$y$ is an input quantum state and 
$\ket{z}$ is
the quantum state produced by the computation $Q(p, y)$, 
and $\ket{x}$ is the target state that one is
trying to describe. 
\end{definition}

\begin{theorem}\label{theo.inv}
There is a universal machine 
\footnote{
We use ``$U$'' to denote a universal (quantum) Turing machine
rather than a unitary matrix.}
$U$ such that for all machines $Q$
there is a constant $c_Q$ (the length of the
description of the index of $Q$ in the enumeration)
such that for all quantum states $\ket{x}$ we have
$K_U (\ket{x} |y) \leq K_Q (\ket{x}|y) + c_Q$. 
\end{theorem}

\begin{proof}
There is a universal quantum Turing machine $U$ in the standard enumeration
$Q_1 , Q_2, \ldots$ such that for every quantum Turing machine
$Q$ in the enumeration there is a self-delimiting program $i_Q$
(the index of $Q$) and $U(i_Q p , y) = Q(p,y)$ for all $p,y$.
Setting $c_Q = l(i_Q)$ proves the theorem.
\end{proof}

We fix once and for all a reference universal quantum Turing machine $U$
and define the quantum Kolmogorov complexity as
\begin{eqnarray*}
&& K (\ket{x} | y) := K_U (\ket{x}|y), \\
&& K (\ket{x}) := K_U (\ket{x} | \epsilon ), 
\end{eqnarray*}
where $\epsilon$ denotes the absence of any conditional
information.
The definition is continuous:
If two quantum states are very close then their quantum Kolmogorov
complexities are very close. Furthermore, since we can approximate
every (pure quantum) state $\ket{x}$ to arbitrary closeness, \cite{BV97},
in particular, for every constant $\epsilon > 0$
we can compute a (pure quantum) state $\ket{z}$
such that 
 $|\bracket{z}{x}|^2 > 1-\epsilon$.
\footnote{We can view this as the probability of the possibly 
noncomputable outcome $\ket{x}$
when executing projection $\shar{x}{x}$ on $\ket{z}$ 
and measuring outcome $\ket{x}$.}
%All outcomes corresponding to vectors
%orthogonal to $\ket{x}$, part of a maximal measurement,
%together have probability $\leq \frac{1}{3}$. If one could perform repeated 
%measurements---possible by repeated effective preparation
%of $\ket{z}$---one can identify outcome $\ket{x}$ among the
%outcomes of this maximum measurement with arbitrary small
%probability of error.
For this definition to be
useful it should satisfy:
\begin{itemize}
\item
The complexity of a pure state that can be directly computed should be the
length of the shortest program that computes that state. (If the
complexity is less then this may lead to discontinuities when we restrict
quantum Kolmogorov complexity to the domain of classical objects.)
\item
The quantum Kolmogorov complexity of a classical object should
equal the classical Kolmogorov complexity of that object (up to
a constant additive term).
\item
The quantum Kolmogorov complexity of a quantum object should
have an upper bound. (This is necessary for the complexity
to be approximable from above, even if the quantum object is
available in as many copies as we require.)
\item
Most objects should be ``incompressible'' in terms of quantum
Kolmogorov complexity.
\item
In a probabilistic ensemble the expected quantum Kolmogorov
complexity should be about equal (or have another meaningful
relation) to the von Neumann entropy.
\footnote{In the classical case the average self-delimiting
 Kolmogorov complexity
equals the Shannon entropy up to an additive constant depending
on the complexity of the distribution concerned.}
\end{itemize}

For a quantum system 
%that certainly passes a test for state 
$\ket{z}$
the quantity $P(x):= |\bracket{z}{x}|^2$ is the probability that
the system passes a test for $\ket{x}$, and vice versa.
The term $\lceil - \log(|\bracket{z}{x}|^2) \rceil$ can be viewed
as the  
code word length to redescribe $\ket{x}$ given $\ket{z}$ 
and an orthonormal basis with $\ket{x}$ as one of the basis vectors
using
the well-known Shannon-Fano prefix code.
This works as follows: For every state $\ket{z}$ in 
$N :=2^n$-dimensional Hilbert space
with basis vectors ${\cal B} = \{ \ket{e_0}, \ldots , \ket{e_{N-1}}\}$ we have
$\sum_{i=0}^{N-1} |\bracket{e_i }{z}|^2 =1$. If the basis has
$\ket{x}$ as one of the basis vectors, then we can
consider $\ket{z}$ as a random variable that assumes value $\ket{x}$
with probability $|\bracket{x}{z}|^2$. The Shannon-Fano code word
for $\ket{x}$ in the probabilistic ensemble
${\cal B}, (|\bracket{e_i}{z}|^2)_i$ is 
based on the probability $|\bracket{x}{z}|^2$ of $\ket{x}$
given $\ket{z}$ and has length  
$\lceil - \log(|\bracket{x}{z}|^2) \rceil$. Considering a canonical
method of constructing an orthonormal basis 
${\cal B} = \ket{e_0}, \ldots, \ket{e_{N-1}}$ 
from a given basis
vector, we can choose ${\cal B}$ such that 
$K({\cal B}) = \min_i \{ K(\ket{e_i}) \} +O(1)$.
The Shannon-Fano code is appropriate for our purpose since it is optimal
in that it achieves the least expected code word
length---the expectation taken over the probability of the
source words---up to 1 bit by Shannon's Noiseless Coding Theorem.

%It is important that the roles of $\ket{x}$ and $\ket{z}$ in
%this definition are not symmetric: $\ket{z}$ is an effectively 
%describable pure quantum state and can possibly be 
%an outcome of a measurement, but $\ket{x}$ can be any one of continuously many
%pure quantum states and therefore be uncomputable (since there
%are only countably many programs and hence computable objects).
%In particular, $\ket{x}$ can be an ``outcome'' that 
%cannot be effectively prepared.

\subsection{Consistency with Classical Complexity}
Our proposal would not be useful if it were the case that for
a directly computable object the complexity is less than the
shortest program to compute that object. This would imply
that the code corresponding to the
probabilistic component in the description is possibly shorter than
the difference in program lengths for programs for an approximation
of the object and the
object itself. This would penalize definite description compared
to probabilistic description and in case of classical objects
would make quantum Kolmogorov complexity less than classical
Kolmogorov complexity. 

\begin{theorem}\label{theo.equiv}
Let $U$ be the reference universal quantum Turing machine
and let $\ket{x}$ be a basis vector in a directly computable orthonormal
basis ${\cal B}$ given $y$: there is
a program $p$ such that $U(p, y)= \ket{x}$.
Then $K(\ket{x} | y)= \min_p \{l(p): U(p, y)= \ket{x} \}$
up to $K({\cal B}|y) +O(1)$.
\end{theorem}
\begin{proof}
Let $\ket{z}$ be such that 
\[
K (\ket{x} | y ) =
\min_{q} \{ l(q) + \lceil - \log(|\bracket{z}{x}|^2) \rceil :
U(q, y) = \ket{z} \} .
\]

Denote the program $q$ that minimizes the righthand side
by $q_{\min}$
and the program  $p$ that minimizes the expression in the statement
of the theorem by $p_{\min}$.

By running $U$ on all binary strings (candidate programs)
 simultaneously dovetailed-fashion
\footnote{A {\em dovetailed} computation is a method related
to Cantor's diagonalization to run all programs alternatingly 
in such a way that every program eventually makes progress. On
an list of programs $p_1, p_2, \ldots$ one divides
the overall computation into stages $k:=1,2, \ldots$. 
In stage $k$ of the overall computation one
executes the $i$th computation step of every program $p_{k-i+1}$
for $i:=1, \ldots , k$.}
one can enumerate all objects that are directly computable given $y$
in order of their halting programs. Assume that $U$ is also
given a $K({\cal B}|y)$ length program $b$ to compute 
${\cal B}$---that is, enumerate the basis
vectors in ${\cal B}$.
This way $q_{\min}$ computes
$\ket{z}$, the program $b$ computes ${\cal B}$.
Now since the vectors of ${\cal B}$ are mutually orthogonal
\[ \sum_{\ket{e} \in {\cal B}} | \bracket{z}{e}|^2 = 1 .
\] 
Since $\ket{x}$ is one of the basis vectors
we have $- \log |\bracket{z}{x}|^2$ is the length of 
a prefix code (the Shannon-Fano code) to compute $\ket{x}$ from $\ket{z}$
and ${\cal B}$.
Denoting this code by $r$ we have that the concatenation $q_{\min} b r$
is a program to compute $\ket{x}$: parse it into 
$q_{\min}, b,$ and $r$ using the self-delimiting 
property of $q_{\min}$ and $b$. Use
$q_{\min}$ to compute $\ket{z}$ and use $b$ to compute ${\cal B}$, 
determine the
probabilities $|\bracket{z}{e}|^2$ for all basis vectors
$\ket{e}$ in ${\cal B}$. Determine the Shannon-Fano code words
for all the basis vectors from these probabilities.
Since $r$ is the code word for $\ket{x}$ we can now
decode $\ket{x}$. Therefore,
\[ l(q_{\min} ) + \lceil - \log(|\bracket{z}{x}|^2) \rceil
\geq l( p_{\min}) - K({\cal B}|y) - O(1) \]
which was what we had to prove.
\end{proof}

\begin{corollary}\label{cor.clasquant}
\rm
On classical objects (that is, the natural numbers
or finite binary strings that are all directly computable) the 
quantum Kolmogorov complexity coincides up
to a fixed additional constant with the self-delimiting
Kolmogorov complexity since $K({\cal B}|n) = O(1)$ for the standard
classical basis ${\cal B}= \{0,1\}^n$.
\footnote{
This proof does not show that it coincide up to an additive constant term
with the original plain complexity defined by Kolmogorov, \cite{LiVi97},
based on Turing machines where the input is delited by distinguished markers.
The same proof for the plain Kolmogorov complexity shows
that it coincides up to a logarithmic additive term. 
}
(We assume that the information about the dimensionality 
of the Hilbert space is given conditionally.)
\end{corollary}

\begin{remark}
\rm
Fixed additional constants are no problem since
the complexity also varies by fixed additional constants due to the choice of
reference universal Turing machine.
\end{remark}

\subsection{Upper Bound on Complexity}
%One way to achieve an *upper bound* of $n$ is to make $p$ a program that just
%outputs $n$ qubits on its "input tape", which specify $\ket{x}$.
%In this case, the "error term" $\log(|\bracket{y}{x}|^2)$
%is zero, so $(|P| - \log(|\bracket{y}{x}|^2)) = n$.
%\begin{corollary}
%By the existence of the identity quantum machine we have:
%For each state $\ket{x}$ consisting of $n$ qubits we have
%$C(\ket{x}) \leq n+O(1)$.
%\end{corollary}
%Another way to achieve an upper bound of $n$ with high probability is to
%make $p$ a very short program that just outputs the state 
%$\ket{ 00 \ldots 0}$.
%Here $l(p)$ can be 1, but the expected value of $ |\bracket{y}{x}|^2$ will be $1/2^n$
%(Wim can explain why). I think that with high probability 
%$|\bracket{y}{x}|^2$
%will be close to its mean, so that with high probability,
%$-\log(|\bracket{x}{y}|^2) = n$. 
%
%There are still other "hybrid" ways of getting the expected upper bound
%to be $n$, such as having $l(p) = n/2 $, and $-\log(|\bracket{y}{x}|^2) = n/2$.

A priori, in the worst case $K(\ket{x} |n )$
is possibly $\infty$. We show that the worst-case has a $2n$ upper bound.

\begin{lemma}
For all $n$-qubit quantum states $\ket{x}$ we
have $K(\ket{x} |n)\leq 2n+O(1)$.
\end{lemma}
\begin{proof}
For every state $\ket{x}$ in $N :=2^n$-dimensional Hilbert space
with basis vectors $\ket{e_0}, \ldots , \ket{e_{N-1}}$ we have
$\sum_{i=0}^{N-1} |\bracket{e_i }{x}|^2 =1$. Hence there is an $i$
such that $|\bracket{e_i }{x}|^2 \geq 1/N$. 
Let $p$ be a $K(i|n)+O(1)$-bit program to construct a
basis state $\ket{e_i}$ given $n$.
Then $l(p) \leq n + O(1)$.
Then $K ( \ket{x} |n ) \leq l(p)  - \log (1/N) \leq 2n + O(1)$.
\end{proof}

\begin{comment}
\begin{remark}
\rm
One may think that $K(\ket{x} ) \leq 3n/2 + O(1)$.
Namely, for every diagonal $d \in \{ d_0, \ldots , d_{2^n - 1} \}$ 
for some basis vector $e$ we have $|\bracket{e}{d}|^2 \geq 1/2^{n/2}$.
Moreover, for every vector $x$ there is a $d$ such that
$| \bracket{e}{x}|^2 \geq | \bracket{e}{d}|^2$.
But the argument seems wrong because there is equal probability
for all basis vectors that a diagonal is observed?
\end{remark}
\end{comment}

\subsection{Computability}

In the classical case Kolmogorov complexity is not computable
but can be approximated from above by a computable process.
The non-cloning property prevents us from copying an unknown pure 
quantum state given to us. Therefore, an approximation from
above that requires checking every output state against the
target state destroys the latter. To overcome the fragility of
the pure quantum target state one has to postulate that it
is available as an outcome in a measurement.

\begin{theorem}
Let $\ket{x}$ be the pure quantum state we want to describe.

{\rm (i)} The quantum Kolmogorov complexity $K(\ket{x})$ is not computable.

{\rm (ii)} 
If we can repeatedly execute the projection $\shar{x}{x}$
and perform a measurement with outcome $\ket{x}$, then
the quantum Kolmogorov complexity $K(\ket{x})$
can be approximated
from above by a computable process with arbitrarily
small probability of error $\alpha$ of giving a too small value.
\end{theorem}
\begin{proof}
The uncomputability follows a fortiori from the classical case.
The semicomputability follows because we have established an upper
bound on the quantum Kolmogorov complexity, and we can simply
enumerate all halting classical programs up to that length by running their
computations dovetailed fashion. The idea is as follows:
Let the target state be $\ket{x}$
of $n$ qubits. Then, $K(\ket{x}|n) \leq 2n + O(1)$. (The 
unconditional case $K(\ket{x})$ is similar 
with $2n$ replaced by $2(n + \log n)$.)
We want to identify a program $x^*$ such that $p:=x^*$ minimizes
$l(p) - \log |\bracket{x}{U(p,n)}|^2$ among all candidate programs.
To identify it in the limit,
for some fixed $k$ satisfying (\ref{eq.alpha}) below 
for given $n, \alpha , \epsilon$,
repeat the computation of every halting program
$p$ with $l(p) \leq 2n+O(1)$ at least $k$ times and perform the assumed
projection and measurement. For every halting program $p$ in the dovetailing
process we estimate the probability 
$q :=|\bracket{x}{U(p,n)}|^2$ from the fraction $m/k$:
the fraction of $m$ positive outcomes out of $k$ measurements.
The probability that the estimate $m/k$ is off from the real
value $q$ by more than
an $\epsilon q$ is given by Chernoff's bound: 
for
$0 \leq \epsilon \leq 1$,
\begin{equation}
\label{chernoff}
P ( |m- qk |  >  \epsilon qk )
\leq 2e^{ - \epsilon^2 qk /3}.
\end{equation}
This means that the probability that the deviation $|m/k - q|$
exceeds $\epsilon q$ vanishes exponentially with growing $k$.
Every candidate program $p$ satisfies
(\ref{chernoff}) with its own $q$ or $1-q$. There are $O(2^{2n})$
candidate programs $p$ and hence also $O(2^{2n})$ outcomes $U(p,n)$
with halting computations. 
%Choose $\epsilon > 0$ so that 
%if $m$ is the number of successes in the experiment
%with probability $q$, and $m'$ that for $q'$, then 
%\begin{equation}\label{eq.q}
%|\frac{m}{k}-\frac{m'}{k}| >  3 \epsilon (\frac{m}{k}+{m'}{k}). 
%\end{equation}
%Then, with probability at most $4e^{ - \epsilon^2 k /3}$ the
%two associated probabilities $q,q'$ do not satisfy
%\begin{equation}\label{eq.q}
%|q-q'| >   \epsilon (q+q') \geq  \epsilon. 
%\end{equation}
We use this estimate to upper bound the probability of error $\alpha$.
For given $k$, the probability
that {\em some} halting candidate program $p$ satisfies
$ |m- qk |  >  \epsilon qk$
is at most $\alpha$ with
\[ \alpha \leq  \sum_{U(p,n) < \infty } 2e^{ - \epsilon^2 q k /3} .\]
The probability that {\em no} halting program does so is
at least $1- \alpha$. That is, with probability
at least $1-\alpha$ we have
\[ (1- \epsilon)q  \leq \frac{m}{k} \leq (1+ \epsilon)q  \]
for every halting program $p$.
It is convenient to restrict attention to the case that all $q$'s are large. 
Without loss of generality,
if $q < \frac{1}{2}$ then consider $1-q$ instead of $q$.
Then, 
\begin{equation}\label{eq.alpha}
 \log \alpha \leq 2n- (\epsilon^2 k \log e )/ 6  +O(1).
\end{equation}

The approximation algorithm is as follows:

{\bf Step 0:} Set the required degree of approximation $\epsilon < 1/2$
and the number of trials $k$ to achieve the required probability of error $\alpha$.

{\bf Step 1:} Dovetail the running of all candidate programs until the
next halting program is enumerated.
Repeat the computation of the new halting program $k$ times

{\bf Step 2:} If there is more than one program $p$ that achieves the
current minimum then choose the program with the smaller length
(and hence least number of successfull observations).
If $p$ is the selected program with $m$ successes out of $k$ trials
then set the current approximation of $K(\ket{x})$ to
\[l(p) - \log \frac{m}{(1+\epsilon)k} .\]
This exceeds the proper value
of the approximation based on the real $q$ instead
of $m/k$ by at most 1 bit for all $\epsilon < 1$.

{\bf Step 3:} Goto {\bf Step 1}.
\end{proof}

\subsection{Incompressibility}

\begin{definition}\label{def.pqscomp}
\rm
A pure quantum state $\ket{x}$ is {\em computable} if 
$K(\ket{x}) < \infty$. Hence all finite-dimensional pure
quantum states are computable. 
We call a pure quantum state {\em directly
computable} if there is a program $p$ such that
$U(p)= \ket{x}$.
\end{definition}

The standard orthonormal basis---consisting of all $n$-bit strings---of
 the $2^n$-dimensional
Hilbert space ${\cal H}_N$ has
at least $2^n (1-2^{-c})$ basis vectors $\ket{e_i}$
that satisfy $K(\ket{e_i} |n) \geq n-c$. This is the standard counting argument
in \cite{LiVi97}. But what about nonclassical orthonormal bases?

\begin{lemma}\label{lem.lowb}
There is a  (possibly nonclassical) orthonormal basis of the $2^n$-dimensional
Hilbert space ${\cal H}_N$ such
that at least $2^n (1-2^{-c})$ basis vectors $\ket{e_i}$ 
satisfy $K(\ket{e_i} |n) \geq n-c$.
\end{lemma}
\begin{proof}
Every orthonormal basis of ${\cal H}_N$ 
has $2^n$ basis vectors and there are at most
$m  \leq \sum_{i=0}^{n-c-1} 2^i = 2^{n-c}-1$ programs of length less than
$n-c$.  Hence there are at most $m$ programs
available to approximate the basis vectors.
We construct an orthonormal basis satisfying the lemma:
The set of directly computed pure quantum states 
$\ket{x_0}, \ldots , \ket{x_{m-1}}$
span an $m'$-dimensional subspace ${\cal A}$ with $m' \leq m$
in the $2^n$-dimensional Hilbert space ${\cal H}_N$ such
that ${\cal H}_N = {\cal A} \oplus {\cal A}^{\perp}$.
Here ${\cal A}^{\perp}$ is a $(2^n - m')$-dimensional
subspace of ${\cal H}_N$ such that every vector in it is
perpendicular to every vector in ${\cal A}$.  
We can write every element $\ket{x} \in {\cal H}_N$ as 
\[
\sum_{i=0}^{m'-1} \alpha_i \ket{a_i}+ \sum_{i=0}^{2^n-m'-1} \beta_i \ket{b_i}
\]
where the $\ket{a_i}$'s form an orthonormal basis
of ${\cal A}$ and the  $\ket{b_i}$'s form an
orthonormal basis of $ {\cal A}^{\perp}$ so that
the $\ket{a_i}$'s and $\ket{b_i}$'s form an orthonormal basis $K$
for ${\cal H}_N$. For every directly computable state
 $\ket{x_j} \in {\cal A}$ and basis vector $\ket{b_i} \in A^{\perp}$ we have
$|\bracket{x_j}{b_i} |^2 = 0$ 
implying that
$K(\ket{x_j}|n) - \log | \bracket{x_j}{b_i} |^2 = \infty$
and therefore $K(\ket{b_i}|n) > n-c$
($0 \leq j < m, 0 \leq i < 2^n - m'$).
This proves the lemma.
\end{proof}

We generalize this lemma to arbitrary bases:
\begin{theorem}\label{theo.lowb}
Every orthonormal basis $\ket{e_0}, \dots ,
\ket{e_{2^n-1}}$ of the $2^n$-dimensional
Hilbert space ${\cal H}_N$  has
at least $2^n (1-2^{-c})$ basis vectors $\ket{e_i}$ 
that satisfy $K(\ket{e_i}|n) \geq n-c$.
\end{theorem}
\begin{proof}
Use the notation of the proof of Lemma~\ref{lem.lowb}.
Assume to the contrary that there are $>2^{n-c}$
basis vectors $\ket{e_i}$ with $K(\ket{e_i}|n) < n-c$.
Then at least two of them, say $\ket{e_0}$
and $\ket{e_1}$ and some
pure quantum state $\ket{x}$ directly computed from a $<(n-c)$-length program
satisfy 
\begin{equation}\label{eq.ex}
K(\ket{e_i}|n) = K(\ket{x}|n) + \lceil - \log |\bracket{e_i}{x}|^2 \rceil .
\end{equation}
($i=0,1$). This means that $K(\ket{x}|n)<n-c-1$ since not both
$\ket{e_0}$ and $\ket{e_1}$ can be equal to $\ket{x}$.
Hence for every directly computed pure quantum state of complexity
$n-c-1$ there is at most one basis state of the same complexity
(in fact only if that basis state is identical with the directly
computed state.)
Now eliminate all directly computed pure quantum states $\ket{x}$ of 
complexity $n-c-1$ together with the basis states $\ket{e}$
that stand in relation Equation~\ref{eq.ex}. We are now
left with $> 2^{n-c-1}$ basis states that stand in relation
of Equation~\ref{eq.ex} with the remaining at most
$2^{n-c-1}-1$ remaining directly computable pure quantum states
of complexity $\leq n-c-2$.
Repeating the same argument we end up with $>1$ basis vector
that stand in relation of Equation~\ref{eq.ex} with 0
directly computable pure quantum states of complexity $\leq 0$
which is impossible.
\end{proof}

\begin{corollary}
The uniform probability 
$\Pr\{\ket{x}: K(\ket{x}|n) \geq n-c \} \geq 1-1/2^c$.
\end{corollary}

\begin{example}
\rm
We elucidate the role of the $- \log  | \bracket{x}{z} |^2$
term.
Let $x$ be a random classical string with $K(x) \geq l(x)$ 
 and let $y$ be a string obtained from $x$
by complementing one bit. It is known (Exercise 2.2.8 in \cite{LiVi97})
that for every such $x$ of length $n$ there is such a $y$ with complexity 
$K(y|n) = n - \log n +O(1)$. Now let $\ket{z}$ be a pure quantum state which has
classical bits except the difference bit between $x$ and $y$ that has
equal probabilities of being observed as ``1'' and as ``0.''
We can prepare $\ket{z}$ by giving $y$ and the position of the
difference bit (in $\log n$ bits) 
and therefore $K(\ket{z}|n) \leq n + O(1)$. Since from
$\ket{z}$ we have probability $\frac{1}{2}$ of obtaining $x$
by observing the particular bit in superposition
and $K(x|n) \geq n$ it follows $K(\ket{z} |n) \geq n + O(1)$ and
therefore $K(\ket{z} |n) = n + O(1)$. 
\begin{comment}
Since $\ket{z}$ is
a directly computable state this is consistent with Corollary~\ref{cor.clasquant}.
\end{comment}
From $\ket{z}$ we have probability $\frac{1}{2}$ of obtaining $y$
by observing the particular bit in superposition which (correctly) yields that
$K(y|n) \leq n +O(1)$.
\end{example}

\subsection{Conditional Complexity}
We have used the conditional complexity $K(\ket{x}|y)$
to mean the minimum sum of the length of a classical program to compute
$\ket{z}$ plus the negative logarithm of the probability of outcome
$\ket{x}$ when executing projection $\shar{x}{x}$ on $\ket{z}$
and measuring, given the pure quantum
state $y$ as input on a separate input tape. 
In the quantum situation the notion of inputs consisting
of pure quantum states is subject to very special rules.

Firstly, if we are given an unknown pure quantum state $\ket{y}$ as
input it can be used only once, that is, it is irrevocably consumed
and lost in the computation. It cannot be copied or cloned without
destroying the original \cite{Pe95}. This phenomenon is
subject to the so-called {\em no-cloning theorem} and means that there is
a profound difference between giving a directly computable pure quantum
state as a classical program or giving it literally. Given as a 
classical program we can prepare and use arbitrarily many copies of it.
Given as an (unknown) pure quantum state in superposition it can be
used as start of a computation only once---unless of course we
deal with an identity computation in which the input state is simply
transported to the output state. This latter computation nonetheless
destroys the input state.

If an unknown state $\ket{y}$ is given as input (in the conditional for example)
then the no-cloning theorem of quantum computing says it can be used
only {\em once}. Thus, for a non-classical pure quantum state $\ket{x}$
 we have
\[ K(\ket{x},\ket{x} | \ket{x}) \leq K(\ket{x})+O(1) \]
rather than
$K(x,x|x)=O(1)$ as in the case for classical objects $x$. This
holds even if $\ket{x}$ is directly computable but is
given in the conditional in the form of an unknown pure quantum state. However,
if $\ket{x}$ is directly computable and
the conditional is a classical program to compute
this directly computable state, then 
that program can be used over and over again.

In the previous example,
if the conditional $\ket{x}$ is directly computable, for example
by a classical program $p$, then we have 
both $K(\ket{x}|p) = O(1)$ and
$ K(\ket{x},\ket{x} | p) = O(1)$.
%Let $\ket{x}^*$ denote a {\em shortest classical program} 
%to compute a directly computable state $\ket{x}$.
In particular, for a classical program $p$ that
computes a directly computable state $\ket{x}$ we have
\[ K(\ket{x},\ket{x} | p) = O(1) .\]

It is important here to notice that a classical program for
computing a directly computable quantum state carries {\em more information}
than the directly computable quantum state itself---much like a
shortest program for a classical object carries more information than the
object itself. In the latter case it consists in partial information
about the halting problem. In the quantum case of a directly
computable pure state we have the additional
information that the state is directly computable {\em and} 
in case of a shortest classical program additional information
about the halting problem.

\subsection{Sub-Additivity}
Quantum Kolmogorov complexity of directly computable pure
quantum states in simple orthonormal bases is {\em sub-additive}:
\begin{lemma}\label{lem.additive}
For directly computable $\ket{x}, \ket{y}$ both of which
belong to (possibly different) orthonormal bases of
Kolmogorov complexity $O(1)$ we have
\[  K(\ket{x}, \ket{y} ) \leq K(\ket{x}|\ket{y}) + K(\ket{y}) \]
up to an additive constant term.
\end{lemma}
\begin{proof}
By Theorem~\ref{theo.equiv} we there is a program $p_y$ to compute $\ket{y}$
with $l(p)= K(\ket{y})$ and a program
$p_{y \rightarrow x}$ to compute $\ket{x}$ from $\ket{y}$ 
with $l(p_{y \rightarrow x}) = K(\ket{x}|\ket{y})$ up
to additional constants. Use $p_y$ to
construct two copies of $\ket{y}$ and $p_{y \rightarrow x}$ to construct
$\ket{x}$ from one of the copies of $\ket{y}$.
The separation between
these concatenated binary programs is taken care of
by the self-delimiting property
of the subprograms. The additional constant term 
takes care of the couple of $O(1)$-bit
programs that are required.
\end{proof}

\begin{comment}
\begin{definition}
\rm
Define the information $I(\ket{x}:\ket{y})$ in pure quantum state $\ket{x}$
about pure quantum state $\ket{y}$ by
$I(\ket{x}: \ket{y}):= K(\ket{x})-K(\ket{x}| \ket{y})$.
\end{definition}
The proof is identical to that of the same relations in the classical
case since we are dealing with directly computable states. The last displayed
equation is known as ``symmetry of information'' since
it states that the information in a classical program for $\ket{x}$ 
about a classical program for $\ket{y}$ is
the same as the information in a classical program
for $\ket{y}$ about a classical program
for $\ket{x}$ up to the
additional logarithmic term.
\[ I(\ket{x}:\ket{y}) = I(\ket{y}:\ket{x}) .\]
\end{comment}

\begin{remark}
\rm
In the classical case we have equality in the theorem (up
to an additive logarithmic term). 
The proof of the remaining inequality, as given in the classical case,
doesn't hold directly for the quantum case. It would require
a decision procedure that establishes equality between 
two pure quantum states without error. While the sub-additivity
property holds in case of directly computable states,
is easy to see that for the general case of pure states
the subadditivity property fails 
due to the ``non-cloning'' property. 
For example for pure states $\ket{x}$ that are not ``clonable'' we
have: 
\[   K(\ket{x}, \ket{x} ) > K(\ket{x}| \ket{x}) + K(\ket{x}) =
K(\ket{x}) + O(1) .\]
\end{remark}
We additionally note:
\begin{lemma}
For all directly computable pure states $\ket{x}$ and $\ket{y}$ we have
$K(\ket{x}, \ket{y}) \leq K(\ket{y}) - \log | \bracket{x}{y}|^2$
up to an additive logarithmic term.
\end{lemma}
\begin{proof}
$ K(\ket{x}|\ket{y}) \leq  - \log | \bracket{x}{y}|^2$ by the proof
of Theorem~\ref{theo.equiv}.
Then, the lemma follows by Lemma~\ref{lem.additive}.
%(Tentative) Suppose $U(p)=\ket{z}, U(q)=\ket{t}$ and
%$K(\ket{x}) = l(p) - \log | \bracket{x}{z}|^2$
%$K(\ket{y}) = l(q) - \log | \bracket{y}{z}|^2$
\end{proof}

\section{Qubit Descriptions}
One way to avoid two-part descriptions as we used above
is to allow qubit programs as input. This leads to the 
following definitions, results, and problems.

\begin{definition}
\rm
The {\em qubit complexity} 
of $\ket{x}$ with respect to quantum Turing machine $Q$
with $y$ as conditional input given for free is
\[
KQ_Q (\ket{x} | y ) :=
\min_{p} \{ l(\ket{p}) :
Q(\ket{p}, y) = \ket{x} \}
\]
where $l(\ket{p})$ is the number of qubits in 
the qubit specification $\ket{p}$,
$\ket{p}$ is an input quantum state,
$y$ is given conditionally, and
$\ket{x}$ is
the quantum state produced by the computation $Q(\ket{p}, y)$:
the target state that one 
describes.
\end{definition}

Note that here too 
there are two possible interpretations for the computation relation
$Q(\ket{p}, y) = \ket{x}$. In the narrow interpretation
we require that $Q$ with $\ket{p}$ on the input tape
and $y$ on the conditional tape halts with $\ket{x}$
on the output tape. In the wide interpretation we require
that for every precision $\delta > 0$ the computation
of $Q$ with $\ket{p}$ on the input tape
and $y$ on the conditional tape
and $\delta$ on a tape where the precision is to be supplied
halts with $\ket{x'}$
on the output tape and $|\bracket{x}{x'}|^2 \geq 1-\delta$. 
Additionally one can require that the approximation finishes
in a certain time, say, polynomial in
$l(\ket{x})$ and $1/\delta$.
In the remainder of this section we can allow either interpretation
(note that the ``narrow'' complexity will always be at least as
large as the ``wide''
complexity).
Fix an enumeration of quantum Turing machines like in Theorem~\ref{theo.inv},
this time with Turing machines that use qubit programs.
Just like before it is now straightforward to derive an Invariance Theorem:
\begin{theorem}
There is a universal machine $U$ such that for all machines $Q$
there is a constant $c$ (the length of a self-delimiting
encoding of the index of $Q$ in the enumeration)
such that for all quantum states $\ket{x}$ we have
$KQ_U (\ket{x} |y) \leq KQ_Q (\ket{x}|y) + c$.
\end{theorem}
We fix once and for all a reference universal quantum Turing machine $U$
and express the {\em qubit quantum Kolmogorov complexity} as
\begin{eqnarray*}
&& KQ (\ket{x} | y) := KQ_U (\ket{x}|y), \\
&& KQ (\ket{x}) := KQ_U (\ket{x} | \epsilon ),
\end{eqnarray*}
where $\epsilon$ indicates the absence of conditional
information (the conditional tape contains the ``quantum state''
with 0 qubits). We now have immediately:
\begin{lemma}
$KQ ( \ket{x} )  \leq l(\ket{x})+O(1)$.
\end{lemma}
\begin{proof}
Give the reference universal machine $\ket{1^n 0} \otimes \ket{x}$
as input where $n$ is the index of the identity quantum Turing machine
that transports the attached pure quantum state $\ket{x}$ to
the output.
\end{proof}

It is possible to define unconditional $KQ$-complexity
in terms of conditional $K$-complexity as follows:
Even for pure quantum states that are not directly computable from
effective descriptions we have
$K( \ket{x} | \ket{x}) = O(1)$. This naturaly gives:

\begin{lemma}
The qubit quantum Kolmogorov complexity of
$\ket{x}$ satisfies
\[ KQ ( \ket{x} ) = \min_{p} 
\{ l( \ket{p}): K(\ket{x} | \ket{p} ) \} +  O(1),\]
where $l(\ket{p})$  denotes the number of qubits in $\ket{p}$.
\end{lemma}
\begin{proof}
Transfer the conditional $\ket{p}$ to the input using an $O(1)$-bit
program.
\end{proof}

We can generalize this definition
to obtain conditional $KQ$-complexity.

\subsection{Potential Problems of Qubit Complexity}
While it is clear that (just as with the previous aproach)
the qubit complexity is not computable, it is unknown to the author
whether one can approximate the qubit complexity from above by
a computable process in any meaningful sense. 
In particular, the dovetailing approach
we used in the first approach now doesn't seem applicable due
to the non-countability of the potentential qubit program candidates.
While it is clear that the qubit complexity of a pure quantum
state is at least 1, why would it need to be more than
one qubit since the probability amplitude can be any complex number?
In case the target pure quantum state is a classical binary string,
as observed by Harry Buhrman,
Holevo's theorem \cite{Pe95} tells us that on average one cannot transmit
more than $n$ bits of classical information by $n$-qubit messages
(without using entangled qubits on the side).
This suggests that for every $n$ there exist classical binary
strings of length $n$ that have qubit complexity at least $n$.
This of course leaves open the case of the non-classical pure quantum 
states---a set of measure one---and
of how to prove incompressibility
of the overwhelming majority of states. These matters have since been
investigated by A. Berthiaume, S. Laplante, and W. van Dam
(paper in preparation). 

\section{Real Descriptions}
A final version of quantum Kolmogorov complexity uses 
computable real parameters to describe the pure quantum state
with complex probability amplitudes.
This requires two reals per complex probability amplitude, that is,
for $n$ qubits one requires $2^{n+1}$ real numbers in the worst case.
Since every computable real number may require a separate program,
a computable $n$ qubit state may require $2^{n+1}$ finite programs.
While this approach does not allow the development of a clean
theory in the sense of the previous approaches, it can be directly
developed in terms of algorithmic thermodynamics---an extension
of Kolmogorov complexity to randomness of infinite sequences
(such as binary expansions of real numbers)
in terms of coarse-graining and sequential Martin-L\"off tests, completely
analogous to Peter G\'acs theory \cite{Ga94,LiVi97}.

\section*{Acknowledgement}
The ideas presented in this paper were developed from 1995 through early 1998.
Other interests prevented me from earlier publication.
I thank Harry Buhrman, Richard Cleve, Wim van Dam, Barbara Terhal, John Tromp,
and Ronald de Wolf for
discussions and comments on QKC.
 
\bibliographystyle{amsplain}

\end{document}